\documentclass[a4paper,twosides]{article}
\usepackage{CJK,multicol,multirow,graphics,fancyhdr,epstopdf}
\usepackage{amsmath,amsfonts,amssymb,bm,upgreek,mathrsfs,ccmap,mathcomp}
\usepackage[pagewise,switch,columnwise]{lineno}
\usepackage[dvipsnames]{xcolor}
\usepackage{CPL-2023}

\usepackage[numbers, sort,compress]{natbib}

\usepackage{tabularx}
\usepackage{comment} 
\usepackage{float}
\usepackage{graphicx}
\usepackage{subfigure}
\usepackage{caption}
\usepackage{hyperref}

\renewcommand{\arraystretch}{1.1} 

\newcommand{\cplyear}{2025} \newcommand{\cplvol}{42}
\newcommand{\cplno}{x} \newcommand{\cplpagenumber}{xxxxxx}
 \newcommand{\cplpage}{\cplpagenumber-\thepage}

\begin{document}
\begin{center}
\large\bf{\boldmath{The Interstellar Scintillation of the Radio-Loud Magnetar XTE~J1810$-$197}}
\footnotetext{\hspace*{-5.4mm}$^{*}$Corresponding authors. Email: yanzhen@shao.ac.cn; zshen@shao.ac.cn

}
\vspace{5mm}\\
\normalsize \rm{}Rui Wang$^{1,2}$, Zhen Yan$^{3,1,2*}$, Zhiqiang Shen$^{3,1,2,4*}$, Zhenlong Liao$^{1,2}$, Zhipeng Huang$^{5}$, Yajun Wu$^{3,1,2}$, Rongbing Zhao$^{3,1,2}$, Xiaowei Wang$^{1,2}$, Jie Liu$^{1}$, Kuo Liu $^{3,1,2}$, Fan Yang$^{1,4}$, Yangyang Lin$^{1}$, Chuyuan Zhang$^{1,2}$ 
\\[3mm]\small\sl $^{1}$Shanghai Astronomical Observatory, Chinese Academy of Sciences, Shanghai, 200030, China;

$^{2}$University of Chinese Academy of Sciences, Beijing, 100049, China;

$^{3}$ State Key Laboratory of Radio Astronomy and Technology, Shanghai Astronomical Observatory, CAS, Shanghai, 200030, China;

$^{4}$School of Physical Science and Technology, ShanghaiTech University, Shanghai, 201210, China;



$^{5}$School of Physics and Mechanical \& Electrical Engineering, Institute of Astronomy and High Energy Physics, Hubei University of Education, Wuhan 430205, China

\vspace{4mm}\normalsize\rm{}(Received xxx; accepted manuscript online xxx)
\end{center}
\vskip 1.5mm

\small{\narrower 
We present a comprehensive interstellar scintillation (ISS) study of the radio-loud magnetar XTE~J1810$-$197, based on six years of multi-frequency monitoring (2018--2024) with the Shanghai Tian Ma Radio Telescope (TMRT) at 7.0, 8.6, and 14.0~GHz. The scintillation parameters—decorrelation bandwidth $\Delta\nu_{\rm d}$, decorrelation time $\Delta\tau_{\rm d}$, and drift rate $dt/d\nu$—are fully characterized. Our measured $\Delta\tau_{\rm d}$ implies $\Delta\tau_{\rm d} < 4$~s at 575--725~MHz under a Kolmogorov spectrum, which is shorter than the magnetar’s 5.54~s spin period. This result naturally explains the previously reported absence of pulse-to-pulse coherence at these frequencies. Kinematic modeling locates the dominant scattering screen at $1.6\pm0.1$~kpc away from the Earth, within the Sagittarius Arm. The screen coincides with the H\,{\sc ii} region JCMTSE~J180921.2$-$201932 and is unrelated to the magnetar’s 2018 outburst suggested by earlier studies. A scintillation arc detected at 14.0~GHz represents the highest-frequency arc observed to date. The asymmetry of arcs is linearly correlated with a dispersion-measure gradient across the screen ($r = 0.959$, $p < 10^{-8}$). We also measure its refractive scintillation timescale, which is only $1.21\pm0.19$~d. Clear DISS at 14~GHz effectively resolves the debate over a possible strong-to-weak scattering transition at this frequency. These results extend the ISS characterization of magnetars to previously unexplored frequencies and provide a precise probe of the ionized interstellar medium in the Sagittarius Arm.
\par}\vskip 3mm
\vskip 5mm

\begin{multicols}{2}
{\it 1. Introduction.} Magnetars are a class of neutron stars distinguished by their ultra-strong magnetic fields ($B\sim10^{14}$–$10^{15}$~G), which are hundreds of times stronger than those of ordinary pulsars. They are characterized by intense, sporadic outbursts of X-rays and soft $\gamma$-rays. The energy released in these outbursts can greatly exceed the rotational energy loss from the star's spin-down, indicating that their emission is powered by the gradual decay and catastrophic restructuring of their immense magnetic fields rather than by rotation \ucite{dt92}. Unlike ordinary radio pulsars, only a small fraction of magnetars have detectable radio signals. Moreover, they exhibit transient emissions distinct from the steady pulsations of ordinary pulsars \ucite{crh16}. So far, only approximately 30 magnetars have been discovered (24 confirmed and 6 candidates), of which only 6 have been detected emitting radio pulsations \ucite{ok14} \footnote{\label{magnetar} \url{https://www.physics.mcgill.ca/~pulsar/magnetar/main.html}}.

XTE~J1810$-$197 (also known as PSR~J1809$-$1943) was discovered in 2003 as an anomalous X-ray pulsar \ucite{ims04}. Subsequent observations identified it as the first radio-loud magnetar with a spin period $P = 5.54$~s and a dispersion measure $DM = \rm 178\pm5~pc~cm^{-3}$ \ucite{hgb05, crh06}. In late 2008, it entered a decade-long quiescent phase before the outburst in both radio and X-ray in December 2018 \ucite{gha19, lld19}. Although the $DM$ of XTE~J1810$-$197 yields distances of 3.6 and 3.1~kpc based on the NE2001 model \ucite{cl02} and the YMW16 model \ucite{ymw17}, respectively, the Very Long Baseline Array (VLBA) parallax measurements provided a closer distance of $2.5\pm0.4$~kpc \ucite{ddl20}. VLBA observations also revealed a proper motion ($\mu_\omega$, $\mu_\delta$) = $(3.70, -16.13)$~$\rm mas~yr^{-1}$ in the barycentric ecliptic coordinate system (corresponding to a transverse velocity $V_{\rm p}\approx200~\rm km~s^{-1}$) and a scattering size of $0.7\pm0.4$~mas at 5.7~GHz. The scattering size of this pulsar indicates a strong interstellar medium (ISM) scattering effect, which is directly associated with interstellar scintillation (ISS), making XTE~J1810$-$197 an ideal target for investigating the ISM through ISS studies. 

Shortly after the discovery of pulsars, diffractive interstellar scintillation (DISS) was proposed to explain flux density fluctuations of pulsars on short timescales ($\sim$~min) and narrow bandwidths ($\sim$~kHz--MHz), characterized by the decorrelation time ($\Delta\tau_{\rm d}$) and the decorrelation bandwidth ($\Delta\nu_{\rm d}$), respectively \ucite{sch68, ric69}. As research progressed, longer-term variations of pulsar flux density (tens of days or more) were also discovered and explained by refractive interstellar scintillation (RISS) \ucite{rcb84}. ISS observations of pulsars serve as a powerful tool for investigating the structure and characteristics of the ISM, such as H\,{\sc ii} regions, supernova remnants, and nebulae \ucite{mma22, yzm21, wys25}. However, very few ISS studies have been conducted on magnetars. Besides the limited number of magnetars with radio pulse radiation, this is mainly due to their erratic radio emissions often masking their tiny ISS patterns. Their relatively long spin period also limits the sub-integration length. Recently, the ISS of XTE~J1810$-$197 at 575--725~MHz was systematically investigated using the upgraded Giant Metrewave Radio Telescope (uGMRT) \ucite{mm24} and the $\Delta\nu_{\rm d}$ of XTE~J1810$-$197 was measured. However, the $\Delta\tau_{\rm d}$ was not acquired, which suggested that measuring $\Delta\tau_{\rm d}$ requires higher frequency observations because $\Delta\tau_{\rm d}$ will be longer and easier to measure in higher frequencies \ucite{Ric77}. The ISS features of XTE~J1810$-$197 were also hypothesized to be associated with the outburst in 2018 and would fade over time; however, no strong evidence was available to test this idea.

In this letter, we present our ISS study results of XTE~J1810$-$197 at 7.00, 8.60, and 14.0~GHz using the Shanghai Tian Ma Radio Telescope (TMRT). The detailed observation information will be provided in Section~2, and the relevant data analysis will be presented in Section~3. In addition to the discussions in Section~4, conclusions will also be presented in the final part.

{\it 2. Observations.} The TMRT is a 65~m fully-steerable antenna built in the 2010s in the Songjiang district of Shanghai City. Equipped with eight sets of low-noise cryogenic receivers and an active surface system, the TMRT can operate over a frequency range from 1.3 to 50.0~GHz. The digital backend system (DIBAS) built for the TMRT consists of three pairs of analog-to-digital converters (ADCs) and three corresponding Roach-II electronic boards \ucite{ysm18}. Throughout the 2018–2024 active phase of XTE~J1810$-$197, we frequently monitored it at 8.60~GHz with the TMRT and published its timing and radiation properties \ucite{hys23}. We selected 83 epochs of data with a long track length, high flux density, minimal radio frequency interference (RFI), and stable pulse profiles from these observations to study its ISS. To investigate its multi-frequency ISS, we also incorporated six epochs of 7.00~GHz and one epoch of 14.0~GHz data with TMRT. The observing bandwidth ($BW$) was 800~MHz at all three frequencies. To mitigate dispersion and RFI, the total $BW$ was divided into sub-channels with typical widths of 1.95~MHz and the typical sub-integration length is 30~s. 

{\it 3. Data Analysis and Results.}
The observational data were pre-processed as follows: First, we applied bandpass calibration using standard flux calibrators (3C~286 or 3C~295). Radio frequency interference (RFI) was then mitigated by excising contaminated sub-integrations and channels interactively using the \texttt{pazi} tool in \textsc{PSRCHIVE} \ucite{hvm04, swd12} \footnote{\label{foot:psr} \url{http://psrchive.sourceforge.net/}}. Subsequently, we generated preliminary dynamic spectra using the \texttt{psrflux} command. To isolate the scintillation signal, intrinsic luminosity variations were suppressed using the \texttt{correct\_dyn} function in \textsc{scintools} \ucite{rc23, rcb20} \footnote{\label{foot:scin}\url{https://github.com/danielreardon/scintools}}, which employs singular value decomposition (SVD). Due to the strong intrinsic variability, the 14.0~GHz data were excluded from this correction step. Finally, any data gaps resulting from RFI mitigation were filled using linear interpolation to ensure continuity in the final dynamic spectra.

The dynamic spectrum plots are employed to directly display how the flux intensity changes with observing frequency and time. Visible scintillation structures (scintles) can be seen at 7.00, 8.60, and 14.0~GHz. Following previous works \ucite{rcn14, rch19, rc23}, we use the two-dimensional (2D) auto-correlation function (ACF) to quantify the average characteristics of scintles ($\Delta\nu_{\rm d}$, $\Delta\tau_{\rm d}$, and the frequency drift rate $dt/d\nu$). The 2D-ACFs were fitted using a 2D exponential model with the open‑source package \texttt{scintools} \ucite{rc23, rcb20} (detailed fitting method in \href{run:supplement.pdf}{Appendix S2} of the online supplementary materials). The DISS strength was quantified by the modulation index, defined as $m = \sigma_{\rm S}/\overline{S(\nu,t)}$, where $\sigma_{\rm S}$ is the standard deviation of flux density $S$ and $\overline{S(\nu,t)}$ is the mean of the flux intensity. In addition, we also use the secondary spectrum, which is the Fourier power spectrum of the pulsar's dynamic spectrum, to investigate the scattering screens in the pulsar's line of sight (LoS) by their parabolic arc structures. To remove the frequency dependence of arc curvature, the dynamic spectrum was firstly re-sampled in the wavelength domain $S(t, \lambda)$ using \textsc{scintools}. The outer 10\% of the data were then tapered with a Hamming window function to mitigate sidelobe effects. Following mean subtraction and pre-whitening, the secondary spectrum was derived via a 2D Fourier transform. To enhance the visibility of arc structures, the logarithmic secondary spectrum $\mathcal{S}$ was obtained from:
\begin{equation}
\mathcal{S}(f_{\rm t},f_{\lambda})=10\log_{10}(|\tilde{S}(t, \lambda)|^2),
\label{equ:ss}
\end{equation}
where $\tilde{S}(t, \lambda)$ is the 2D Fourier transform of the pre-processed wavelength-domain dynamic spectrum. The scintillation arcs are parameterized by their curvature $\eta$ and half-width $w$, with the arc regions defined by the inequality $|f_{\lambda} - \eta f_{\rm t}^2| < w$. In this expression, $f_{\rm t}$ and $f_{\lambda}$ denote the Fourier conjugates of the time $t$ and wavelength $\lambda$ axes, respectively. Parameters $\eta$ and $w$ were determined using a Hough-transform-based fitting method \ucite{xlh18}. 

\begin{figure*}[htb]
\vspace{0.1cm}
\flushleft
	\subfigure{\begin{minipage}[b]{0.98\linewidth}
    \centering
\includegraphics[height=12cm,width=14cm]{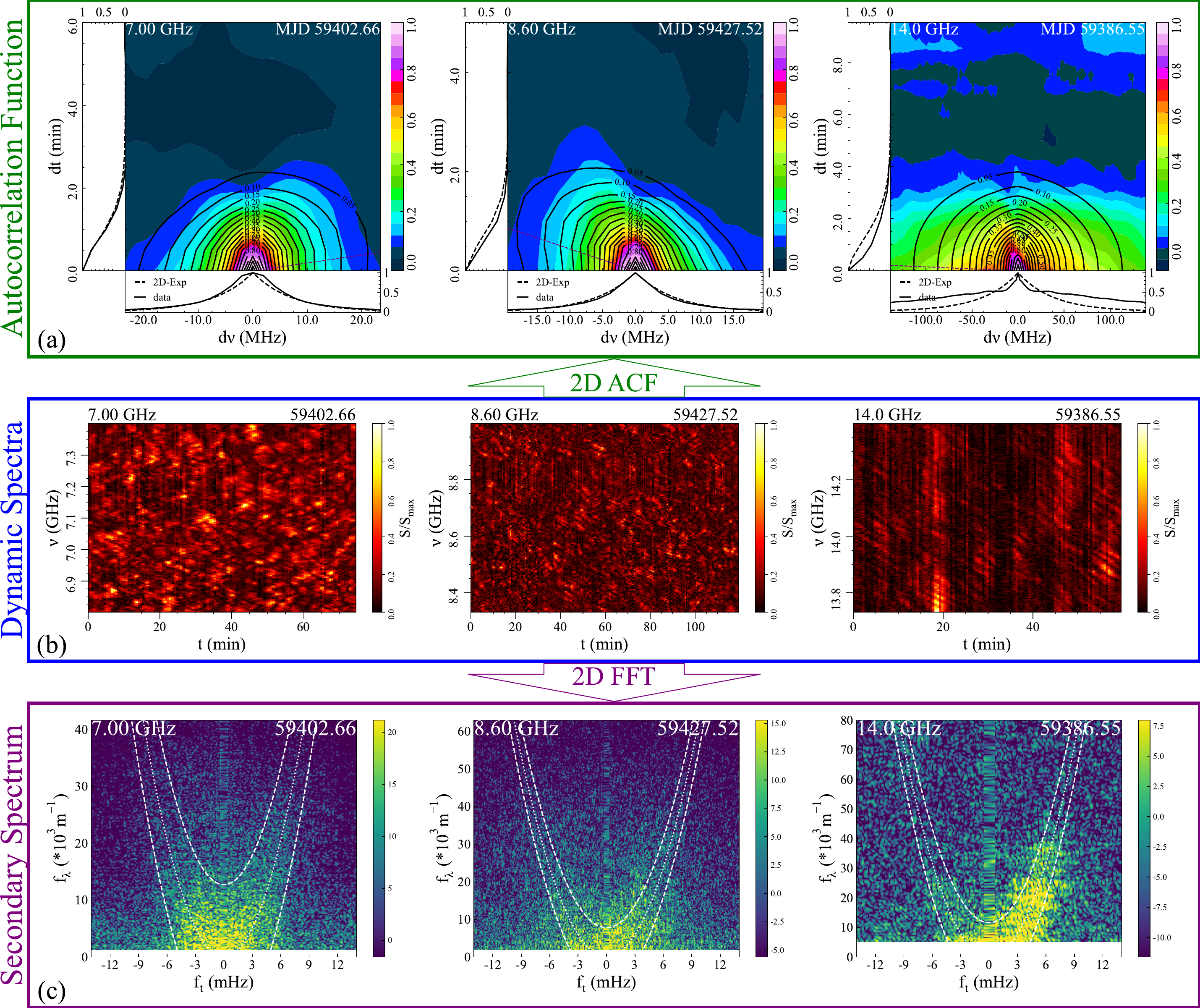}  
\end{minipage}}
\vspace{-0.25cm}
\caption{Representative plots of ISS data analysis results for observations listed in Table~\ref{tab:ourresults}. (a) 2D-ACFs: The normalized 2D-ACFs are shown with color-filled contours (scaled with the color-bar on the right), and the 2D-Exponential form fitting results are black-line contours. The major axis is marked with the dashed line. The bottom and left sub-panels are the 1D cuts of the 2D-ACF (solid curves) and the best-fit model (dashed curves) at zero time and frequency lag. (b) Dynamic spectra: The intensity of each pixel is linearly scaled with the color-bar on the right. (c) Secondary spectra: The white dashed curves and white dotted curves denote the boundaries and center of the best-fit arc regions, respectively. }
\vspace{-0.5cm}
\label{fig:opra}

\end{figure*}

At 8.60~GHz, we found that $\Delta\tau_{\rm d}$ ranged between 0.85 and 1.80~min, with a mean of 1.22~min, only about twice the time resolution of our data. $\Delta\nu_{\rm d}$ ranged from 4.2 to 13.5~MHz, averaging 8.1~MHz. The DISS modulation index $m$ ranges irregularly between 0.49 and 1.38, so the ISS of this magnetar remains in the strong regime. These parameters show no systematic evolution over time. The values of $\Delta\tau_{\rm d}$ and $\Delta\nu_{\rm d}$ for the 7.00~GHz observation are smaller than those for 8.60~GHz, while $m$ is similar. The only 14.0~GHz observation is influenced by strong intrinsic variability. Detailed discussions are presented in Section 4. Representative dynamic spectra, 2D-ACFs, and secondary spectra for each observation frequency are presented in Figure~\ref{fig:opra}; the corresponding ISS analysis results are listed in Table~\ref{tab:ourresults}. The complete set of plots and the detailed data table for all observations are provided in the online supplementary materials (\href{run:supplement.pdf}{Appendix S1–S3} and \href{run:supplement.pdf}{Table S1}, respectively).

\begin{table*}
\caption{The information and results of representative our ISS observations at each frequency: observation frequency ($\nu$), Modified Julian Date (MJD), observation length ($T$), decorrelation bandwidth ($\Delta\nu_{\rm d}$), decorrelation time ($\Delta\tau_{\rm d}$), drift rate ($dt/d\nu$), curvature of arc ($\eta$), half-width of arc ($w$), logarithmic asymmetry of arc ($\ln({S_{\rm r}}/{S_{\rm l}})$), reduced ISS speed ($V_{\rm iss, 0}$), $DM$ gradient (${\partial DM}/{\partial t}$), DISS modulation index ($m$) and their 1-$\sigma$ uncertainties (if measured). Complete table can be seen in \href{run:supplement.pdf}{Table S1} in the online supplementary materials.}
\vspace{-0.25cm}
\fontsize{8}{11}\selectfont
\setlength{\tabcolsep}{1.75pt}
\label{tab:ourresults}
\renewcommand{\arraystretch}{1.3}
\begin{tabularx}{\linewidth}{ccccccccc ccc c}
\hline\hline
$\nu$ &\multirow{2}{*}{MJD} &  $T$ & $\Delta\nu_{\rm d}$ & $\Delta\tau_{\rm d}$ & $dt/d\nu$ & $\eta$ & $w$ & \multirow{2}{*}{$\ln({S_{\rm r}}/{S_{\rm l}})$} & $V_{\rm iss,0}$ & ${\partial DM}/{\partial t}$ & \multirow{2}{*}{$m$} & \\
(GHz) && (s) & (MHz) & (min) & (s/MHz) & ($\rm m^{-1}mHz^{-2}$) & ($\rm 10^3~m^{-1}$) & & ($\rm km~s^{-1}$) & ($\rm pc~cm^{-3}yr^{-1}$) & \\
\hline
7.00 &59402.65& 4480 & $4.95\pm0.12$ & $1.22\pm0.03$ & $1.15\pm0.07$ & $573_{-29}^{+42}$ & $12.7_{-2.8}^{+1.1}$ & $-0.12_{-0.01}^{+0.01}$ & $261\pm22$ & $-0.05\pm0.01$ & $0.67$ \\
\hline
8.60 &59427.50& 7155 & $4.23\pm0.07$ & $1.01\pm0.02$ & $-2.66\pm0.08$ & $661_{-25}^{+55}$ & $7.8_{-1.5}^{+2.3}$ & $0.35_{-0.03}^{+0.01}$ & $238\pm20$ & $0.16\pm0.01$ & $0.77$ \\
\hline
14.0 &59386.52& 3578 & $27.96\pm1.90$ & $1.97\pm0.13$ & $-0.09\pm0.03$ & $881_{-22}^{+100}$ & $11.6_{-3.0}^{+1.2}$ & $1.48_{-0.03}^{+0.04}$ & $194\pm21$ & $0.02\pm0.01$ & $0.92$ \\
\hline\hline
\end{tabularx}
\vspace{-0.5cm}
\end{table*}

{\it4. Discussions.}

{\it4.1 Comparison with previous studies.}
\citet{mm24} conducted an ISS observation of XTE~J1810$-$197 using the uGMRT previously. They measured $\Delta\nu_{\rm d}$ in the 575--725~MHz band, and it was predicted that $\Delta\tau_{\rm d}$ would be 6.9--9.4~s at these frequencies according to the NE2001 model \ucite{cl02}; however, an attempt to constrain it from the cross-correlation of pulses failed. Extrapolating our measurements of $\Delta\tau_{\rm d}$ to 575--725~MHz (assuming it follows a Kolmogorov spectrum $\Delta\tau_{\rm d} \propto \nu^{1.2}$ \ucite{Ric77}) gives $\Delta\tau_{\rm d} < 4$~s, which is shorter than the 5.54~s spin period, thereby accounting for the absence of pulse-to-pulse coherence at those frequencies. They also examined the frequency dependence of $\Delta\nu_{\rm d}$, which follows $\Delta\nu_{\rm d} \propto \nu^{\alpha}$ \ucite{Ric77}. Their analysis yielded a best-fit $\alpha = 4.0\pm0.3$, but a Kolmogorov spectrum ($\alpha = 4.4$) was also consistent with their data; they consequently concluded consistency with Kolmogorov turbulence. Extrapolating their result to 8.60~GHz yields $\Delta\nu_{\rm d} \approx 11.9\pm0.6$~MHz (for $\alpha=4.4$) or $4.23\pm0.22$~MHz (for $\alpha=4.0$). Both extrapolations are consistent with our direct measurement from 4.2 to 13.5~MHz. This paper also hypothesized that the ISS was associated with the 2018 outburst and would diminish over time; however, our six-year monitoring campaign found no such evolution.

In contrast to the prediction of a strong-to-weak ISS transition at 14.6~GHz with $\Delta\nu_{\rm d}\simeq 320$~MHz \ucite{ljk08}, the extrapolation of uGMRT observations yields $\Delta\nu_{\rm d}\simeq 32$~MHz at 14~GHz — a value nearly one order of magnitude smaller \ucite{mm24}. We have only one ISS observation at 14.0~GHz (MJD~59386.52). The analysis was conducted on uncorrected data due to strong intrinsic variability, which appears as vertical stripes in the dynamic spectrum and a ridge at $\Delta\tau\approx0$ in the 2D-ACF (Figure~\ref{fig:opra}). Consequently, the derived $\Delta\nu_{\rm d} = 27.96\pm1.90$~MHz may be overestimated. Nevertheless, this value agrees with the extrapolation of uGMRT observations \ucite{mm24} and is significantly lower than the theoretical prediction \ucite{ljk08}, challenging the predicted strong-to-weak transition scenario. This conclusion is further supported by $m \approx 1$ at 14.0~GHz, which indicates a strong ISS at this frequency.

\begin{figure*}[hbt]
\subfigure{\begin{minipage}{1\linewidth}
\centering
\includegraphics[height=4.8cm,width=12cm]{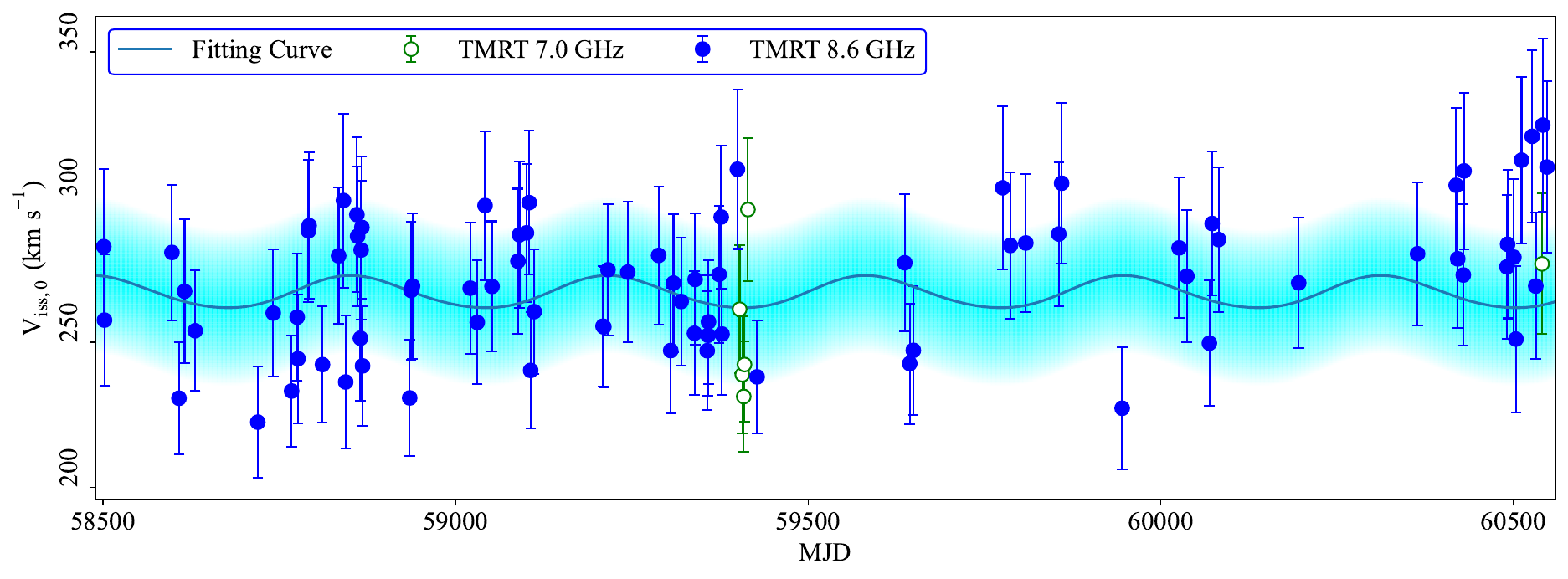}       
\end{minipage}}
\vspace{-0.5cm}
\caption{Observed $V_{\rm iss, 0}$ and the fitting curve. $V_{\rm iss, 0}$ at 7.0 and 8.6~GHz and their 1-$\sigma$ error ranges are marked with hollow green and solid blue points with error-bars. The blue curve with a shaded region is the fitting curve and its 1-$\sigma$ error ranges.} 
\vspace{-0.5cm}
\label{fig:viss}
\end{figure*}

{\it 4.2 Location of the scattering screen.}
The ISS speed $V_{\rm iss}$ is defined as the velocity of the diffraction pattern relative to the observer. For a thin screen at a distance of $L_{\rm o}$ from the observer and $L_{\rm p}$ from the pulsar, with a total distance of $D = L_{\rm o} + L_{\rm p}$, $V_{\rm iss}$ is given by:
\begin{equation}
    V_{\rm iss} = A_{0} \frac{\sqrt{\Delta\nu_{\rm d} D x}}{\nu \Delta\tau_{\rm d}} = \sqrt{x} V_{\rm iss,0},
    \label{eq:vissbasic}
\end{equation}
where $x=L_{\rm o}/L_{\rm p}$ is the fractional screen distance. The quantity $V_{\rm iss,0} = A_{0} \sqrt{\Delta\nu_{\rm d} D} / (\nu \Delta\tau_{\rm d})$ denotes the reduced ISS speed. The model-dependent constant $A_{0} = 3.85 \times 10^{4}$ \ucite{Gup95} yields $V_{\rm iss}$ in $\rm km~s^{-1}$ when the input parameters $\Delta\nu_{\rm d}$, $D$, $\nu$, and $\Delta\tau_{\rm d}$ are in units of MHz, kpc, GHz, and seconds, respectively.

ISS velocity $\mathbf{V}_{\rm iss}$ is modeled as a weighted vector combination of the pulsar's transverse proper motion ($\mathbf{V}_{\rm p, \bot}$, where $\bot$ denotes the component perpendicular to the LoS), the Earth's orbital velocity ($\mathbf{V}_{\rm earth, \bot}$), and the transverse velocity of the scattering screen ($\mathbf{V}_{\rm scr, \bot}$):
\begin{equation}
\mathbf{V}_{\rm iss} = x\mathbf{V}_{\rm p, \bot} + \mathbf{V}_{\rm earth, \bot} - (1+x)\mathbf{V}_{\rm scr, \bot}.
\label{eq:vissvec}
\end{equation}
For an isolated pulsar, $\mathbf{V}_{\rm p, \bot}$ and $\mathbf{V}_{\rm scr, \bot}$ can be treated as constants over multi-year timescales \ucite{cr98, xsl23}.

The observed $V_{\rm iss,0}$ for XTE~J1810$-$197 showed substantial scatter, which prevented a complete fit. We therefore adopted the common simplification of a stationary scattering screen ($\mathbf{V}_{\rm scr, \bot}=0$), an approach justified when the pulsar's transverse speed greatly exceeds that of the screen \ucite{Gup95}. Under this assumption, for a solitary pulsar at coordinates ($\omega_{\rm p}, \delta_{\rm p}$) with a proper motion ($\mu_\omega, \mu_\delta$) in $\rm mas~yr^{-1}$ in the barycentric ecliptic coordinates, there is \ucite{bpl02, wys25}:
\begin{equation}
\begin{aligned}
xV_{\rm iss,0}^2 &= [4.74xD\mu_\omega - V_{\rm e}\cos(\omega_{\rm e}-\omega_{\rm p})]^2 \\&+
[4.74xD\mu_\delta - V_{\rm e}\sin{\delta_{\rm p}}\sin(\omega_{\rm e}-\omega_{\rm p})]^2.
\end{aligned}
\label{visseclip3}
\end{equation}
where $V_{\rm e}$ and $\omega_{\rm e}$ are the Earth's orbital speed and ecliptic longitude. A least-squares fit to $V_{\rm iss,0}$ yielded $x = 1.8\pm0.4$. This fitting result indicates that the variation of $V_{\rm iss,0}$ from the revolution of the Earth is smaller than 11~km~$s^{-1}$, compared to the measurement uncertainties of $V_{\rm iss,0}$, while $\mathbf{V}_{\rm p, \bot}$ is about 200~km~$s^{-1}$. Consequently, $V_{\rm iss,0}$ is dominated by $V_{\rm p, \bot}$, resulting in a less prominent variation trend of $V_{\rm iss,0}$. Figure~\ref{fig:viss} displays an imperfect fit curve, but it matches most $V_{\rm iss,0}$ within the error range, so $x = 1.8\pm0.4$ is acceptable. Using the VLBA parallax distance of $D=2.5\pm0.4$~kpc \ucite{ddl20}, $x = 1.8\pm0.4$ corresponds to $L_{\rm o}= 1.6\pm0.1$~kpc and $L_{\rm p}= 0.9\pm0.1$~kpc. 

The derived $L_{\rm o} = 1.6\pm0.1$~kpc places the scattering screen within the Sagittarius Arm, far from the magnetar but near the foreground H\,{\sc ii} region JCMTSE~J180921.2$-$201932. This H\,{\sc ii} region, located at a distance of $d_{\rm H} = 1.56$~kpc and with a projected separation of $d_{\rm H,\bot} \approx 16$~pc from the LoS to XTE~J1810$-$197 \ucite{csu13}, is an active high-mass star formation region \ucite{ygm23}. Its localized ionization would lead to a higher local electron density than traditional Galactic electron density models predict, offering a natural explanation for why both the NE2001 \ucite{cl02} and YMW16 \ucite{ymw17} models overestimate the distance to XTE~J1810$-$197. This location also indicates that the observed ISS is dominated by the ISM along the LoS rather than the local environment of the magnetar, excluding any association with the 2018 outburst. Furthermore, our assumption of a stationary screen is consistent with the Galactic differential rotation model \ucite{mua19}, which predicts a transverse speed $<0.2$~km~s$^{-1}$ of this screen.
 
{\it4.3 Curvature of arc.}
The curvature $\eta$ of the parabolic arc in the secondary spectrum is determined by the screen geometry and the velocity of the point in the screen intersected by the LoS to the pulsar $\mathbf{V}_{\rm eff, \bot}$. According to the anisotropic thin-screen model \ucite{crs06}:
\begin{equation}
\eta = \frac{D x / (1+x)^2}{2 (\mathbf{V}_{\rm eff, \bot} \cdot \mathbf{a})^2},
\label{equ:etaani}
\end{equation}
where the unit vector $\mathbf{a}$ signifies the orientation of the scattering image's major axis, which is the sky projection of the underlying turbulence anisotropy \ucite{Shi21}. Isotropic scattering occurs in the specific case where $\mathbf{a}$ is always parallel to $\mathbf{V}_{\rm eff, \bot}$.

The effective transverse velocity, $\mathbf{V}_{\rm eff, \bot}$, is also a weighted combination of the transverse velocities of the Earth, the pulsar, and the scattering screen \ucite{smc01, crs06}:
\begin{equation}
\mathbf{V}_{\rm eff, \bot} = \frac{x}{1+x} \mathbf{V}_{\rm p, \bot} + \frac{1}{1+x} \mathbf{V}_{\rm earth, \bot} - \mathbf{V}_{\rm scr, \bot}.
\label{equ:v}
\end{equation}
We project $\mathbf{V}_{\rm eff, \bot}$ onto the ecliptic coordinate system, yielding components ${V}_{\rm eff, \omega}$ (ecliptic longitude) and ${V}_{\rm eff, \delta}$ (ecliptic latitude). Let $\psi_{0}$ be the position angle of the scattering image's major axis measured clockwise from the ecliptic latitude direction, $\mathbf{a} = (\cos\psi_{0}, \sin\psi_{0})$. Substituting this into Equation~\eqref{equ:etaani} gives \ucite{wys25}: 
\begin{equation}
\eta = \frac{D x / (1+x)^{2}}{2 (V_{\rm eff,\omega} \cos\psi_{0} + V_{\rm eff,\delta} \sin\psi_{0})^{2}}.
\label{equ:eta2}
\end{equation}

\begin{figure*}[htbp]
\subfigure{\begin{minipage}{1\linewidth}  
\centering
\includegraphics[height=4cm,width=12cm]{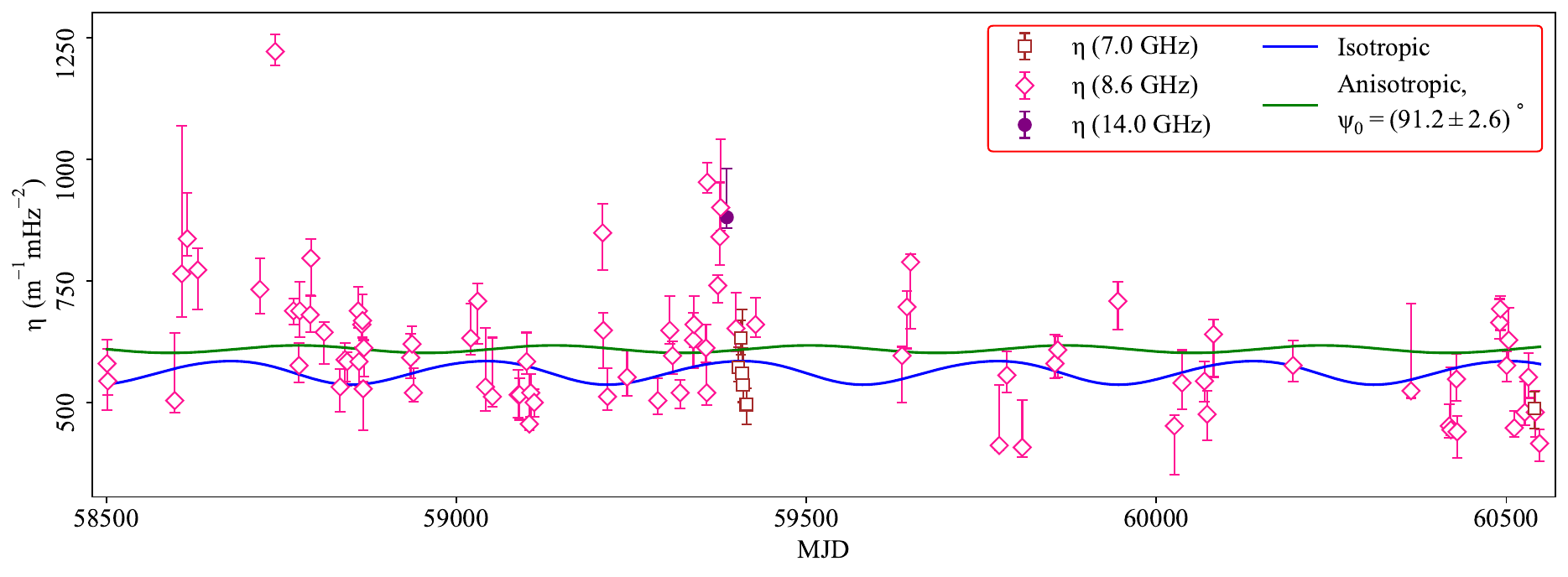}       
\end{minipage}}
\vspace{-0.5cm}
\caption{The $\eta$ values changing with the observation time. $\eta$ at 7.00, 8.60, and 14.0~GHz and their 1-$\sigma$ error ranges are hollow brown squares, hollow deeppink diamonds, and solid purple points with asymmetric error-bars, respectively. The best-fit curves for isotropic (blue) and anisotropic (green) scattering scenarios are also shown on the plots.}
\vspace{-0.25cm}
\label{fig:eta}
\end{figure*}

Figure~\ref{fig:eta} presents the best-fit curves for both isotropic (blue) and anisotropic (green) scattering models, overlaid on the measured $\eta$ from our 7.00 (yellow diamonds), 8.60 (pink diamonds), and 14.0~GHz (purple diamond) observations. We can see that both models provide comparable fits. This degeneracy originates from the screen geometry and velocity dominance: with $x = 1.8\pm0.4$ and $V_{\rm p, \bot} \gg V_{\rm earth, \bot}$, the effective velocity $\mathbf{V}_{\rm eff}$ is dominated by the pulsar's motion. Consequently, $\mathbf{V}_{\rm eff}$ remains nearly constant in both magnitude (variations $<4\%$) and direction (deviating $<8^\circ$ from $\mathbf{V}_{\rm p, \bot}$). Under these conditions, anisotropic scattering behaves similarly to the isotropic case when the major axis is aligned with $\mathbf{V_{\rm p, \bot}}$. Our anisotropic fit yields a small misalignment ($<15^\circ$) between the major axis [$\psi_0=(91.2\pm2.6)^\circ$] and $\mathbf{V_{\rm p, \bot}}$ ($\psi_{\rm _{V_{p}}} \approx 102^\circ$, where $\psi_{\rm _{V_{p}}}$ is the position angle of the line of $\mathbf{V_{\rm p, \bot}}$ measured clockwise from the ecliptic latitude direction). We conclude that, for our current precision in $\eta$ measurements, the difference between isotropic and anisotropic scattering is drowned in noise, showing no prominent trends for us to distinguish them.

{\it4.4 Asymmetric arcs.}
The parabolic arcs in our secondary spectra exhibit significant asymmetry. This feature can arise from a $DM$ gradient across the scattering screen \ucite{crs06, rcn14} or from a physically finite screen extent \ucite{Shi21}. We quantified this asymmetry using the logarithmic arc asymmetry ratio, defined as $\ln(S_{\rm r}/S_{\rm l})$, where $S_{\rm r}$ and $S_{\rm l}$ are the integrated powers in the right ($f_{\rm t} > 0$) and left ($f_{\rm t} < 0$) halves of the arc region ($|f_{\lambda} - \eta f_{\rm t}^{2}| < w$), respectively. 

To identify the physical origin of the arc asymmetry, we calculated the dispersion measure gradient along the direction of $\mathbf{V}_{\rm eff,\bot}$, $\partial DM/\partial t$,using the following relation \ucite{rc23} to determine whether the $DM$ gradient is associated with the asymmetry of the arcs:
\begin{equation}
\frac{\partial DM}{\partial t} = -\left( \frac{dt}{d\nu} \right) \left( \frac{\pi\nu^3 V_{\rm iss, 0}^{2}}{c^2 D r_{\rm e}} \right),
\label{eq:thetar2}
\end{equation}
where $r_{\rm e}$ is the classical electron radius and $c$ is the speed of light. 

\begin{figure*}[htbp]
\subfigure{\begin{minipage}{0.98\linewidth} 
\centering
\includegraphics[height=5.6cm,width=12.7cm]{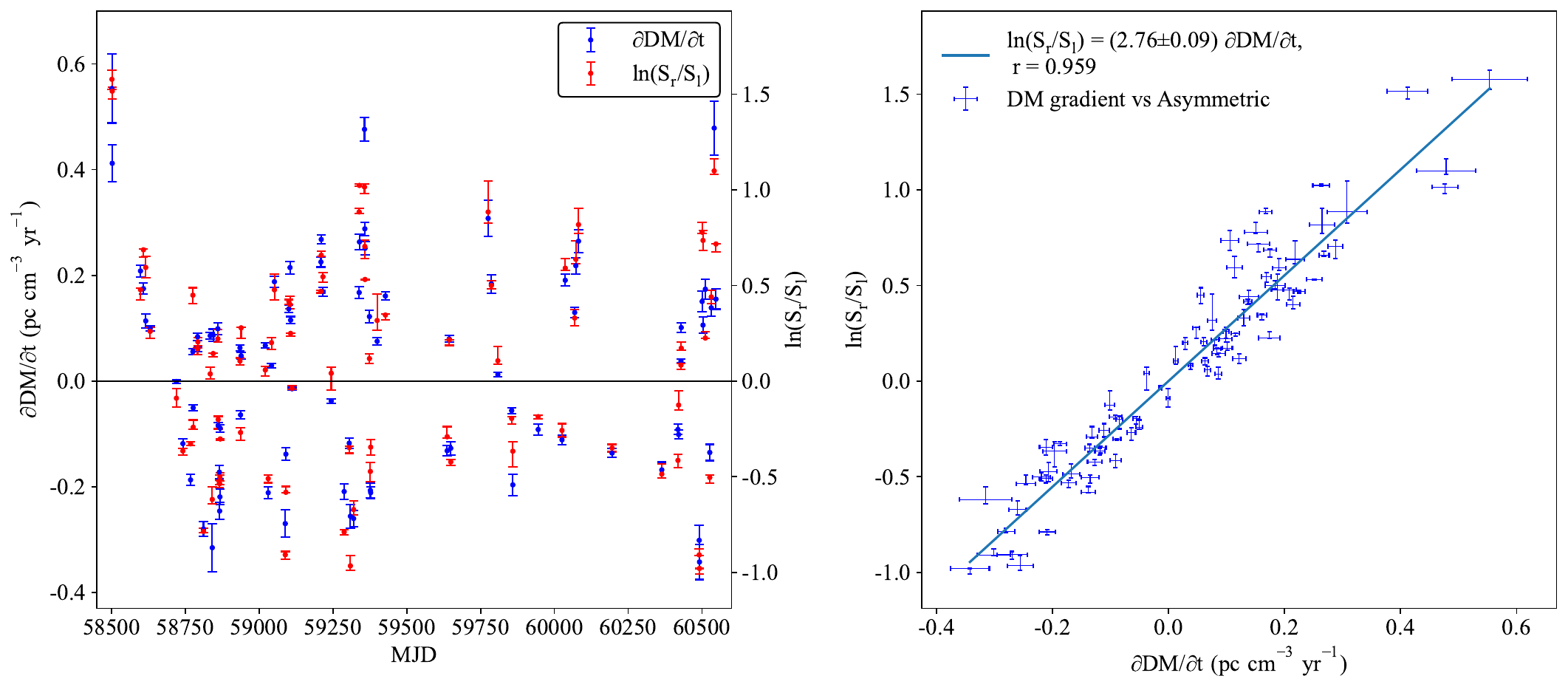}       
\end{minipage}}
\vspace{-0.5cm}
\caption{$\ln({S_{\rm r}}/{S_{\rm l}})$ and ${\partial DM}/{\partial t}$ of XTE~J1810$-$197 at 8.60~GHz. The left panel displayed how ${\partial DM}/{\partial t}$ and $\ln({S_{\rm r}}/{S_{\rm l}})$ varied with time. $\ln({S_{\rm r}}/{S_{\rm l}})$, ${\partial DM}/{\partial t}$ and their 1-$\sigma$ error ranges are marked with red and blue points with error-bars, respectively. The right panel displays the linear relationship between $\ln({S_{\rm r}}/{S_{\rm l}})$ and ${\partial DM}/{\partial t}$. We use crossing error-bars to mark the 1-$\sigma$ error ranges} of $\ln({S_{\rm r}}/{S_{\rm l}})$ and ${\partial DM}/{\partial t}$.
\label{fig:asy}
\vspace{-0.5cm}
\end{figure*}

The left panel of Figure~\ref{fig:asy} shows a clear synchronous evolution of ${\partial DM}/{\partial t}$ and $\ln(S_{\rm r}/S_{\rm l})$ at 8.60~GHz. The observed variation of $\partial DM/\partial t$ across epochs indicates inhomogeneities within the scattering screen on small scales. The right panel of Figure~\ref{fig:asy} displays a highly significant linear correlation between $\partial DM/\partial t$ and $\ln(S_{\rm r}/S_{\rm l})$: $\ln(S_{\rm r}/S_{\rm l}) = (2.76\pm0.09)\cdot{\partial DM}/{\partial t}$, Pearson correlation coefficient $r = 0.959$, $p < 10^{-8}$. This robust relationship strongly suggests a causal connection between the $DM$ gradient and the arc asymmetry. We therefore favor the interpretation that the $DM$ gradient within the scattering screen is the dominant cause of the observed arc asymmetry. In the future, a more refined structural analysis could be conducted using Very Long Baseline Interferometry (VLBI) ISS observations \ucite{ysj24}.

\begin{figure*}[hbt]
\subfigure{\begin{minipage}{0.48\linewidth}
\flushright
\includegraphics[height=4cm,width=6cm]{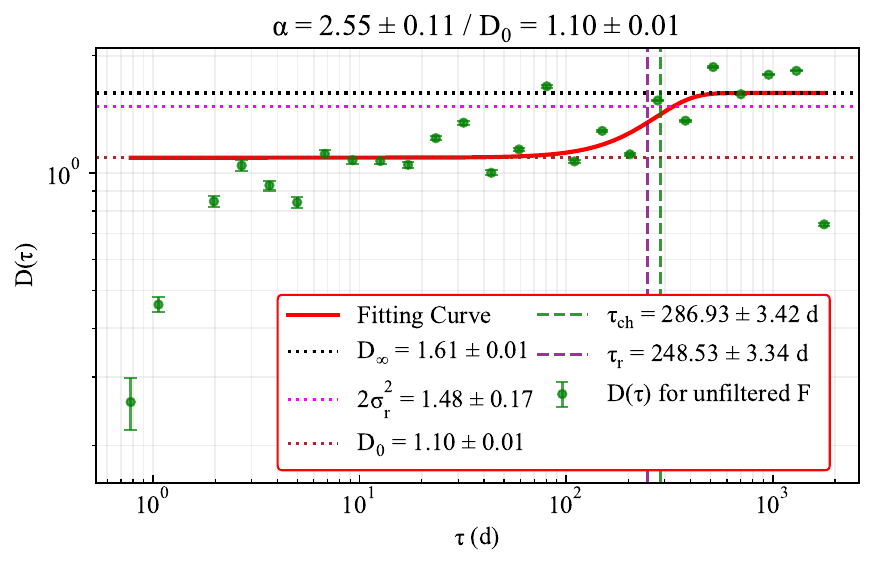}       
\end{minipage}}
\subfigure{\begin{minipage}{0.48\linewidth}
\flushleft
\includegraphics[height=4cm,width=6cm]{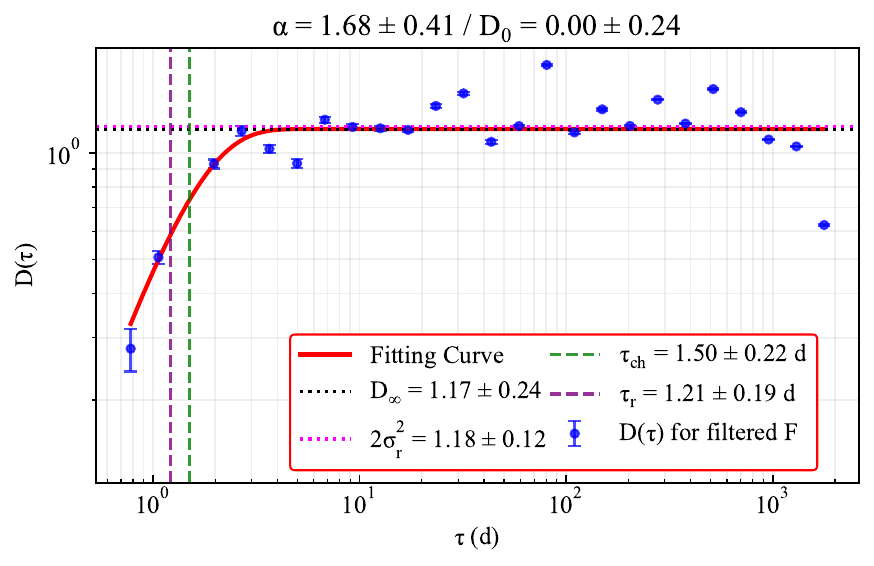}       
\end{minipage}}
\vspace{-0.5cm}
\caption{$D(\tau)$ and the fitting results. $D(\tau)$ and their 1-$\sigma$ error ranges for unfiltered and filtered $F$ are plotted by green and blue dots with error-bars in the left and right panels, respectively. The red curve is the fitting result. Green and purple dashed lines are the $\tau_{\rm ch}$ and $\tau_{\rm r}$, respectively. Black and magenta dotted lines are $D_{\infty}$ and $2\sigma_{\rm r}^{2}$, respectively, where $\sigma_{\rm r}$ is the RISS modulation index. For a reliable fitting, $D_{\infty}\approx 2\sigma_{\rm r}^{2}$. The brown dotted line is $D_{0}$. When $D_{0}$ is close to zero, we would not plot it.} 
\vspace{-0.5cm}
\label{fig:riss}
\end{figure*}

{\it4.5 Refractive scintillation.}
Long‑term flux density ($F$) variations of XTE~J1810$–$197 \ucite{hys23} likely include the RISS component, so we also analyzed the RISS behavior using the 8.6~GHz monitoring data. The theoretical RISS timescale $\tau_{\rm r}$ can be estimated by $\tau_{\rm r} = \nu \Delta\tau_{\rm d}/{\Delta\nu_{\rm d}}$. Using the average $\Delta\tau_{\rm d}$ and $\Delta\nu_{\rm d}$ obtained from our measurements, we find $\tau_{\rm r}\approx 0.90$~d \ucite{lk04}. We calculated and fitted the structure function $D(\tau)$ \ucite{gbs23} to quantify the RISS properties by saturation level $D_{\infty}$, structure index $\alpha_{\rm r}$, characteristic timescale $\tau_{\rm ch}$, noise level $D_0$, and $\tau_{\rm r} = (\ln 2)^{1/\alpha_{\rm r}} \tau_{\rm ch}$ \ucite{ks92, rl90}. However, the long-term intrinsic $F$ evolution of the magnetar severely affects $D(\tau)$, leading to a poor fit (left panel of Figure~\ref{fig:riss}). To suppress slow trends, we applied a moving‑average filter with a window length of 250~d (the $\tau_{\rm r}$ estimated from the unfiltered $F$). After filtering, the fit improves significantly and gives $\tau_{\rm r}\approx1.21\pm0.19$~d (right panel of Figure~\ref{fig:riss}), which is slightly larger than the theoretical estimate of 0.90~d. The complete procedure for calculating and fitting $D(\tau)$ is detailed in the online supplementary material \href{run:supplement.pdf}{Appendix S4}. We note that $\tau_{\rm r}$ is close to the shortest cadence between our observations, which may lead to an overestimation. 

{\it5. Conclusion.} 
Through a six-year multi-frequency monitoring campaign (December 2018–September 2024), we performed the first comprehensive analysis of the ISS from the radio-loud magnetar XTE~J1810$-$197. By combining TMRT observations at 7.00, 8.60, and 14.0~GHz, we obtained a complete set of ISS parameters—$\Delta\nu_{\rm d}$, $\Delta\tau_{\rm d}$, and $dt/d\nu$—fully characterize its ISS behavior. Extrapolating our measurements to the 575–725~MHz band yields $\Delta\tau_{\rm d} < 4$~s, shorter than the spin period ($P = 5.54$~s), thereby explaining the absence of pulse-to-pulse correlations \ucite{mm24}. Kinematic modeling locates the scattering screen in the Sagittarius Arm at $L_{\rm o} = 1.6 \pm 0.1$~kpc, close to the H\,{\sc ii} region JCMTSE~J180921.2$-$201932. This screen rules out the association between the observed ISS and the magnetar’s 2018 outburst. We detected a scintillation arc at 14.0~GHz — the highest-frequency arc observed to date, surpassing the previous record of 8.60~GHz \ucite{wys25}. A strong linear correlation between $\ln(S_{\rm r}/S_{\rm l})$ and $\partial DM/\partial t$ ($r = 0.959$, $p < 10^{-8}$) confirms that the $DM$ gradient across the screen is the primary cause of arc asymmetry. We also estimate its $\tau_{\rm r} = 1.21\pm0.19$~d at 8.60~GHz.

Our study significantly expands the ISS characterization of XTE~J1810$-$197 to higher frequencies but finds no evidence of the predicted strong-to-weak scattering transition up to 14.0~GHz \ucite{ljk08}. Future observations above 15~GHz may directly determine the transition frequency. Moreover, the discrepancy between the VLBA parallax distance and the $DM$-derived distance indicates that current Galactic electron density models systematically underestimate electron densities along the LoS of XTE~J1810$-$197.

\textit{Acknowledgments.}
This work was supported in part by the National SKA Program of China (Grant No.~2020SKA0120104), the National Key R~$\&$~D Program of China (Grant No. 2022YFA1603104), and the National Natural Science Foundation of China (Grant Nos. U2031119, 12041301). 

\newcommand{\nat}{Nature}
\newcommand{\apj}{Astrophys. J.}
\newcommand{\apjl}{Astrophys. J. Lett.}
\newcommand{\apjs}{Astrophys. J. Suppl. S}
\newcommand{\mnras} {Mon. Not. R. Astron. Soc.}
\newcommand{\pasa} {Publ. Astron. Soc. Aust.}
\newcommand{\aap} {Astronom. Astrophys.}
\newcommand{\aj} {Astron. J.} 
\newcommand{\araa} {Annu. Rev. Astron. Astrophys.}
\newcommand{\ptr}{Phil. Trans. R. Soc. Lond. Ser. A}
\newcommand{\chjas}{Chin. J. Astronom. Astrophys. Suppl.}
\newcommand{\art}{Astronom. Res. Technol.}
\newcommand{\pasp}{Publ. Astron. Soc. Pac.}
\newcommand{\jgrsp}{J. Geophys. Res.: Space Phys.}
\newcommand{\ajp}{Aust. J. Phys.}
\newcommand{\scpma}{Sci. China Phys. Mech. Astron.}

\bibliographystyle{iopart-cpl}
\setlength{\bibsep}{0pt}
\bibliography{J1809}

\end{multicols}
\end{document}